\documentclass[onecolumn,letter]{IEEEtran}
\usepackage{times}
\usepackage{graphicx}
\usepackage{amsmath}
\usepackage{amsfonts}
\usepackage{amssymb}
\usepackage{nicefrac}
\usepackage{psfrag}
\usepackage{multirow}

\newtheorem{theorem}{Theorem}

\title{Markov Modeling of Cooperative Multiplayer Coupon Collectors' Problems}

\author{Riccardo Rovatti, Cristiano Passerini, Gianluca Mazzini}

\date{\today}

\newcommand{\goal}{{\rm\bf goal}}

\newcommand{\this}{{\rm\bf this}}
\newcommand{\first}{{\rm\bf first}}

\newcommand{\eoc}{{\pmb{\it eoc}}\;}
\newcommand{\coc}{{\pmb{\it coc}}\;}
\newcommand{\Exp}{\mathbb{E}}
\newcommand{\EBgoal}{\mathcal{B}^{\goal}}
\newcommand{\EBthis}{\mathcal{B}^{\this}}
\newcommand{\EBfirst}{\mathcal{B}^{\first}}
\newcommand{\EOC}{\mathcal{C}^{\rm off}}
\newcommand{\ETC}{\mathcal{C}^{\rm tra}}
\newcommand{\ERC}{\mathcal{C}^{\rm req}}

\newcommand{\pstart}{\begin{pmatrix}1 & 0 & \dots & 0\end{pmatrix}}
\newcommand{\pcollecttop}{\begin{pmatrix}1 & \dots & 1 \end{pmatrix}}

\begin{document}

\maketitle

\begin{abstract}

The paper introduces a modified version of the classical Coupon
Collector's Problem entailing exchanges and cooperation between
multiple players. Results of the development show that, within a
proper Markov framework, the complexity of the Cooperative Multiplayer
Coupon Collectors' Problem can be attacked with an eye to the modeling
of resource harvesting and sharing within the context of Next
Generation Network. The cost of cooperation is computed in terms of
exchange protocol burden and found to be dependent on only ensemble
parameters such as the number of players and the number of coupons but
not on the detailed collection statistics. The benefits of
cooperation are quantified in terms of reduction of the average number
of actions before collection completion.
\end{abstract}

\section{Introduction}

The classical Coupon Collector's Game is a process in which a player
randomly receives coupons corresponding to one out of $M$ possible
labels and continues playing until she collects a coupon for each
possible label (see, e.g., \cite{Rosen_1970}).

Coupon Collector's Problems (CCPs) deal with the statistical
characterization of what happens in a Coupon Collector's Game and,
even in their simplest form, are the object of many classical and
recent investigations. The reason for this interest is twofold. On one
side they offer challenging questions in the field of mathematical
statistics (see, e.g.,
\cite{Papanicolau_1998}\cite{Adler_2001}\cite{Adler_2003}\cite{Myers_2004}\cite{Neal_2008}).
On the other side they enjoy a plethora of applications mostly derived
from general Bayesian capture-recapture method \cite{Robert_2001}
specialized to the most diverse fields such as: counting of biological
species or phenomena (see, e.g. \cite{McCready_2001} and
\cite{Poon_2005}), IP tracking and Internet characterization (see,
e.g., \cite{Ma_2006} and \cite{Kundu_2009}), recognition of the author
of a textual documents \cite{Robert_2001} as well as many other
information processing/managing tasks (see, e.g.,
\cite{Flajolet_1992}, \cite{Boneh_1997}, \cite{Bach_1998}, and
\cite{Deb_2006}).

Despite this large availability of theoretical and practical results,
little is present in the Literature about CCPs with more than one
player (see \cite{Adler_2003} and \cite{Neal_2008} for some early
treatment). In particular, what is never addressed is the case in which
all players aim at completing their collection while accepting to
exchange coupons with other players.

Yet, Cooperative Multiplayer CCPs (CM-CCPs) are promising models for
the behavior and performance of resource managing mechanisms in
nowadays and foreseeable information and communication technology
systems, such as resource harvesting and sharing in Next Generation
Networks (NGN) but also knowledge dissemination in social
infrastructure as well as content distribution in peer-to-peer
organizations to name a few.

Note, in fact, that NGN will have to cope with the problem of
using resources in un-known, un-planned and un-supervised
environments. Such an environment can be coped with by means of a
resource management that hinges on two key steps: resources harvesting
and resource negotiation/exchanging. Though resource harvesting can be
a trivial, even randomized task. Negotiation and exchanging needs
intelligence and cooperation. Intelligence is needed to administer the
trade-off between resources acquisition and release, and balance local
and global profit, present and perspective gains. Cooperation is
needed to release a resource when this can be reasonably expected to
favor global performance.

Within that general framework, consider the example of the
distribution of DRM-protected multimedia content to a group of $P$
subscribers. The complete content is made of individually protected
and distinguishable pieces, that can be unlocked by only one (not
pre-determined) final user. The provider pushes the pieces (possibly
packed in envelopes containing more than one piece) into an anycast
routing network that delivers them to subscribers. Due to unknown
internal network mechanisms users receive random pieces at random
times, retain and unlock what they miss to complete the content, and
offer other pieces to other users through a multicast routing. Other
subscribers request and possibly obtain offered pieces by unicast
routing. Note that the network used by the content provider is not
necessarily the same as the one used for content exchange between
players, thus leading to a noteworthy increase in service deployment
flexibility.

The overall system is made of a content provider, one of more intelligent
distribution networks, and a group of active users whose behavior can
be modeled as a CM-CCP.

Note that CM-CCPs that model effective and efficient distribution
mechanisms must be such that each coupon entering the system is
either assigned to a player who needs it or discarded if all the
players already have it. This is to say
either that at any instant the number of players that still need a
certain coupon is equal to the number of players minus the number of
times the coupon entered the system, or that the events {\em ``the
  coupon this label enters the system a number of times
  equal to the number of players''} and {\em ``all the players have
  the coupon with this label''} are equivalent.

As a final remark, it is sensible to expect that cooperation plays an
important role in distinguishing CM-CCP with $P>1$ players from $P$
non-interacting instances of CCP. First, as a matter of fact,
interaction entails a cost that must be accounted for. Second, even
limited cooperation yields advantages in the achievement of players'
goal. Hence, classical results on single-player CCPs would only yield
upper bounds on time-to-completion performance figures.

\section{Problem statement}

To formally define the problem assume that $P$ players collect
coupons, each of them labeled with of one out of $M$ different
labels. The local goal of each player is to acquire one coupon for
each possible label. The common goal of the players is to ensure that
everybody achieves its local goal as soon as possible.

An activity burst is triggered by the availability of a lot of $L$
coupons having different labels. The incoming lots are drawn
independently. When a lot enters the system it is assigned to only one
player, who is selected independently each time, and so that each
player has the same probability of being assigned the lot.

Once a player receives a lot, she retains the coupons she misses and
offers the remaining (duplicate) ones to the other players. Then, a
contention phase ensues in which each offered coupon is randomly
assigned to one of the players requesting it, each of them having the
same probability of being assigned an offered coupon. A final transfer
phase closes activities, in which assigned coupons are actually
transferred between offering and assigned players. All and only
requested coupons are transferred. No new burst is started once the
common goal has been achieved.

\begin{figure}[t!]
\begin{center}
\psfragscanon
\psfrag{P0}[cc][cc][0.6]{$0$-th}
\psfrag{P1}[cc][cc][0.6]{$1$-th}
\psfrag{P2}[cc][cc][0.6]{$2$-th}
\psfrag{P3}[cc][cc][0.6]{$(P-2)$-th}
\psfrag{P4}[cc][cc][0.6]{$(P-1)$-th}
\includegraphics[width=\columnwidth]{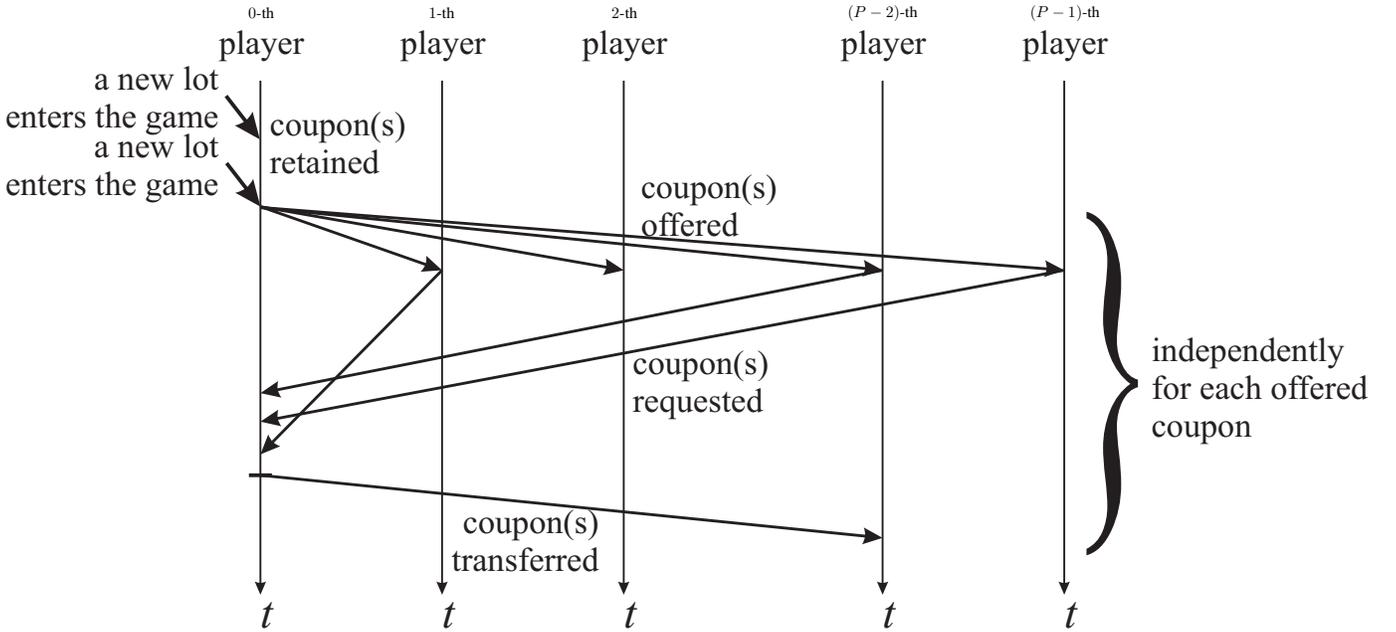}
\psfragscanoff
\end{center}
\caption{\label{protocol}Protocol for coupon's exchanging.}
\end{figure}

In Figure \ref{protocol} we schematize the steps of the protocol
adopted for coupon's exchange. When the coupon fits in the collection
of the player first receiving it, no action is initiated. On the
contrary, when a coupon can be exchanged, the three-phases protocol
with offer, request and transfer is adopted. Transfers occur after all
requests have been collected.

Other protocols may be devised to regulate coupon exchange
between players. For example, though no hint of a true cooperative
multiplayer structure is present, the mechanism in \cite{Adler_2003}
can be extrapolated to define a priority-based exchange.

Comparison between different protocol options in terms of the many
merit figures that may characterize them (overhead, scalability,
flexibility, fairness, etc.) is out of the scope of this paper, which
does not aim at protocol optimization but at introducing cooperation
in CCP and analyzing its effects. From this point of view the reasons
to choose the simple protocol above are at least threefold:
\begin{itemize}

\item it is completely symmetric and thus avoids unfair
  behaviors a priori;

\item involved entities (the one assigning the lot at the beginning of
  the burst and the players exchanging coupons) need a very small
  amount of information on the game structure and state: this avoids
  any significant startup phase (whose cost should be accounted for)
  and easily copes with a varying number of active players;

\item such simplicity does not prevent it from capturing the costs and
  benefits of cooperation.
\end{itemize}

If all players remain in activity until the common goal is achieved
even if they have achieved their local goal we indicate it as a
"continue-on-completion" (\coc) game. If, on the contrary, players
that have completed their collections exit the game, we indicate it as
an "exit-on-completion" (\eoc) game. The two alternative mechanisms
are trivially equivalent for $P=1$ while they are, in principle,
different for $P>1$. Results will be given for \coc
games. Relationships between \coc and \eoc games will be developed to
extend these results to the latter.

The most intuitive quantification of the effectiveness of cooperation
mechanisms is the average number of bursts needed to reach the common
goal, that we will indicate with $\EBgoal$. Concentrating on a
specific player, another interesting quantity is the average number of
bursts needed to complete her collection that we will indicate with
$\EBthis$. Finally, it may be of interest to know how many bursts are
needed, on average, for the first player to complete her collection, a
performance figure that we will indicate as $\EBfirst$.

Beyond this performance figures, three further quantities may be taken
into account as cost factors, namely: the average number of offered
coupons ($\EOC$) that accounts for the traffic due to the offering
phase, the average number of requested coupons ($\ERC$) that accounts
for the traffic due to the response of the players to the offering
phase, and the average number of transferred coupons ($\ETC$) that
accounts for the traffic actually needed to make cooperation
advantageous for players that receive coupons they would have not been
assigned if they were playing alone.

\section{Costs}

As far as the costs of \coc and \eoc schemes are concerned, we may
prove the following

\begin{theorem}
\label{theeocltcoc}

The costs $\EOC$, $\ERC$, and $\ETC$ of a \eoc game are smaller than
the same costs of an \coc game with the same $P>1$, $M$, and $L$.

\end{theorem}

\begin{IEEEproof}
At any given time, no matter whether in a \coc or in a \eoc game,
$P'\le P$ players have not yet finished their collection.
When a new coupon enters the system $P''\le P'$ players miss it in
their collection and are ready to compete for it, if it is offered.

The probability that it is offered is $\pi_{\rm \eoc}=(P'-P'')/P'$ for
an \eoc game and $\pi_{\rm \coc}=(P-P'')/P$ for a \coc game.
From $P'\le P$ we get $\pi_{\rm \eoc}\le \pi_{\rm \coc}$ and thus that
$\EOC$ is smaller for an \eoc game.

Note also that each time a coupon is offered, it triggers a number of
requests equal to $P''$. In expectation this implies that also $\ERC$
is smaller for an \eoc game.

Finally, each time a coupon is offered, it triggers one transfer if
$P''>0$ and no transfer if $P''=0$. In expectation this implies that
also $\ETC$ is smaller in an \eoc game.
\end{IEEEproof}

In the following discussion we will derive expressions for the costs
in the \coc game that Theorem \ref{theeocltcoc} guarantees to be upper
bounds for the \eoc case.

\begin{theorem}
\label{theETC}
Independently of the statistics of lot drawing, the
average number of transferred coupons is
\[
\ETC=\frac{M}{2}(P-1)
\]
\end{theorem}

\begin{IEEEproof}
Up to $P$ coupons of the same type can enter the system and cause a
transfer. Assume to sample the game at each of the corresponding $P$
time instants.

The $j$-th time a coupon of a chosen type enters the system, $j-1$
players already have it in their collection, hence there is a
probability $(j-1)/P$ that the coupon is assigned to one of these
players that will initiate the exchange and eventually produce a
transfer.

Therefore, the average number of transfers of coupons of that type is
$\sum_{j=1}^{P} (j-1)/P=(P-1)/2$.

Since the histories of the $M$ different types of coupon are
independent, the total average number of transferred coupons is
$\frac{M}{2}(P-1)$.
\end{IEEEproof}

\begin{theorem}
\label{theERC}

Independently of the statistics of lot drawing, the
average number of requested coupons is
\[
\ERC=\frac{M}{6}\left(P^2-1\right)
\]

\end{theorem}

\begin{IEEEproof}
As in the proof of Theorem \ref{theETC} let us sample the game at
every entrance of a coupon of a chosen type.

The $j$-th time that this happens, $j-1$
players already have it in their collection and $P-j+1$ do not.

Hence, there is a probability $(j-1)/P$ that the coupon is offered to
other players thus producing $P-j+1$ requests.

Therefore, the average number of transfers of coupons of that type is
$\sum_{j=1}^{P} (P-j+1)(j-1)/P=(P^2-1)/6$.

Since the histories of the $M$ different types of coupon are
independent, the total average number of transferred coupons is
$\frac{M}{6}(P^2-1)$.
\end{IEEEproof}

\begin{theorem}
\label{theEOC}
The average number of offered coupons is
\[
\EOC=\EBgoal L -\frac{M}{2}(P+1)
\]
\end{theorem}

\begin{IEEEproof}
Each offered coupon is either transferred (if a player exists missing
it in her collection) or discarded (if no player needs it).

The average number of transferred coupons is, from Theorem
\ref{theETC}, $\frac{M}{2}(P-1)$.

As far as the number of discarded coupons is concerned, note that,
since $\EBgoal$ is the average number of bursts needed for all the
players to complete their collection, the average number of coupons
entering the game is $\EBgoal L$. The total average number of
discarded coupons is the number of coupons entering the system minus
the $MP$ coupons fitting in the collections.

Putting all together
\[
\EOC=\frac{M}{2}(P-1)+\EBgoal L - MP=\EBgoal L -\frac{M}{2}(P+1)
\]

Note that the last expression can be also interpreted as the total
average number of coupons entering the game minus the average number
of coupons received directly by the players needing them, that
results to be $\frac{M}{2}(P+1)$.
\end{IEEEproof}

\subsection{Remarks}

The above Theorems highlight some properties of the costs
associated to the cooperation mechanism.

First, $\ETC$ and $\ERC$ are independent of the probability that a
coupon with a specific label enters the game. Their value do not
change even if different labels appear with different
probabilities. Hence, these two quantities are strictly linked to the
cooperation mechanism per se.

Actually, this could have been anticipated thinking that requests and
transfers depend only on the entrances of coupons that are needed by
at least one of the players. 
Since all types of coupons will eventually enter the system, their
probabilities cannot affect $\ERC$ and $\ETC$.

The same can be said of $L$, since the player first receiving the lot
treats each coupon separately.

Another feature that is common to $\ETC$ and $\ERC$ is the linear
dependence on $M$: doubling the number of labels implies doubling
these cooperation costs.

As far as the dependency on $P$ is concerned, note that though the
cost due to transfers is proportional to the number of players, the
effort implied by signalling coupon requests increases quadratically
with it since all players needing a coupon apply to obtain it but only
one of them is selected to receive it.

Note finally, that by now no conclusive result on $\EOC$ is given due
to its dependency on $\EBgoal$ that will be evaluated in the next
Sections.

\section{Performance}

In this Section we will assume that all lots are drawn uniformly,
i.e., so that each of the possible $\binom{M}{L}$ lots has the same
probability.

Note that since they refer to instant at which either only one or all
the players have finished, $\EBgoal$ and $\EBfirst$ of an \eoc game
are equal to the same performance figures of a \coc game with the same
$P$, $M$, and $L$.

When $L=1$, a much stronger equivalence holds between the statistical
characterization of the whole game evolution in the \coc and \eoc
case. In fact

\begin{theorem}
\label{theeoceqcoc2}

If $L=1$ the probability that the coupon entering the system at the
beginning of a burst is finally retained by a player pre-selected among
those that need it in their collection is the same for a \coc and an
\eoc game with the same $P$ and $M$.

\end{theorem}

\begin{IEEEproof}
At any given time, no matter whether in a \coc or in a \eoc game,
$P'\le P$ players have not yet finished their collection.
When a new coupon enters the system $P''\le P'$ players miss it in
their collection and are ready to compete for it, if it is offered.

Assuming to chose one of these $P''$ players, the probability that she
receives the new coupon is
\[
\frac{1}{P}+\frac{P-P''}{P}\frac{1}{P''}=\frac{1}{P''}
\]
\noindent for a \coc game, and
\[
\frac{1}{P'}+\frac{P'-P''}{P'}\frac{1}{P''}=\frac{1}{P''}
\]
\noindent for an \eoc game. 
\end{IEEEproof}

Note that the above strict equivalence cannot hold in general for
$L>1$. As a counterexample think of a system with $P=3$ players, $M=2$
different labels and $L=2$ coupons per lot. The average number of
bursts needed by the player that is the second to complete her
collection can be easily computed in the \eoc and \coc case.

In a \eoc game, the first player that receives a lot completes her
collection and exits, hence $\EBfirst=1$. Between the two remaining
players, the one receiving the second lot is the second to complete
her collection. Hence, the number of bursts needed to complete the
second collection is 2.  Only one player remains that completes her
collection as soon as the third lot enters the system, thus
$\EBgoal=3$.

In a \coc game, the first player that receives a lot completes her
collection, hence $\EBfirst=1$ as expected.

The second lot entering the system can be assigned either to one of
the two players with no coupon or to the player that has completed the
collection. In this latter case (that holds with probability
$\frac{1}{3}$), the receiving player assigns the two coupons
independently to the other players with uniform probability. The
probability that the coupons are finally assigned to two different
players is $\frac{1}{4}+\frac{1}{4}=\frac{1}{2}$. Overall, the
probability that no further player completes her collection at the
second burst is $\frac{1}{3}\frac{1}{2}=\frac{1}{6}>0$ thus implying
that the average number of bursts needed by the player that is the
second to complete her collection is strictly greater than 2.

Regardless whether a further player has completed her collection in
the second burst, once a third lot comes in the coupons are surely
distributed to complete all remaining collection and the game finishes
thus making $\EBgoal=3$ as expected.

\subsection{Markov model}

The aim of this section is to count the number of bursts needed for
game completion. Hence, we need to track the game evolution from its
initial state toward the achievement of the final common goal.

A straightforward option would be to evolve a vector containing the
number of times a certain label enters the system, i.e., an $M$-tuple
of integers taking values from $0$ to $P$ and thus assuming one of
$(P+1)^M$ different values.

Note that, since all labels are statistically indistinguishable, what
really matters is not how many times each individual labels appeared
during the game. Rather, we may group labels that appeared once,
twice, thrice, and so on, and simply count the cardinality of each of
these subsets.

Since each of the first $P$ times a coupon with a certain label enters
the systems it is assigned to a player that needs it, we may
characterize the system by recording how many of the $M$ labels are
such that a certain number of players has a corresponding coupon.

More formally, we define a integer $(P+1)$-tuple $S=(S_P,\dots,S_0)$
where $S_j$ is the number of labels for which exactly $j$ players have
a coupon. Clearly
\begin{equation}
\label{Sdef}
\begin{array}{l}
S_j\ge 0\\
\sum_{j=0}^P S_j=M
\end{array}
\end{equation}

\noindent so that one of the components contains a redundant
information.

Due to the sum constraint, the number $\mu_P(M)$ of distinct
$P$-tuples corresponding to observable states is
equal to the number of ways in which the integer $M$ can be
decomposed as the sum of $P+1$ non-negative integers or, equivalently,
the number of ways in which one can choose $P$ objects among $M+P$,
i.e.,
\[
\mu_P(M)=\binom{M+P}{P}
\]

Counting with $n$ the number of bursts since the beginning of the
game, we have $S(0)=(0,\dots,0,M)$. After that, $S(n)$ can be any of
the integer vectors satisfying \eqref{Sdef}. In particular, if
$S(n)=(M,0,\dots,0)=\goal$ then the players have reached their common
goal and the game is over.

Since the lots are drawn independently of the state currently
characterizing the system, the evolution does not depend on the past
states but on the current one. Hence, the transitions probabilities
\begin{equation}
\label{eq:TS}
T_{S'',S'}=\Pr\{S(n+1)=S''|S(n)=S'\}
\end{equation}

\noindent given for any possible $S',S''\neq\goal$ are sufficient to
describe the overall game evolution.

Assuming that $S(\bar{n})=\goal$ then, the sequence of states $S(n)$
for $n=0,\dots,\bar{n}-1$ is a non-stationary stochastic process whose
first-order characterization is given by the probabilities
$p_{S'}(n)=\Pr\{S(n)=S'\}$ for each possible $S'\neq\goal$. Using the
probabilities \eqref{eq:TS} we have that
\[
T_{S'',S'}\;p_{S'}(n)
\]

\noindent is the joint probability that $S(n+1)=S''\neq\goal$ and
$S(n)=S'\neq\goal$ and that
\[
p_{S''}(n+1)=\sum_{S'} T_{S'',S'}\;p_{S'}(n)
\]

\noindent is the relationship between the first-order characterization
of the system after $n$ and after $n+1$ bursts.

As a final remark on our choice of the state characterizing the system
note the following

\begin{theorem}
\label{theScoceoc}

If a starting state $S'$ is given, along with an incoming lot, then
the system state $S''$ at the end of the burst is the same in a \coc
and in an \eoc game.

\end{theorem}

\begin{IEEEproof}
Among the $L$ coupons in the lot, $L'\le L$ coupons are missed by at
least one player. Regardless of the game mechanism, those $L'$
coupons will be eventually assigned to one of those players
incrementing by one the number of player that own them. Since the state
$S''$ will take into account the number of players owning each type of
coupon, it will be the same for \coc and \eoc games.
\end{IEEEproof}

Seen from the point of view of the state $S$, the dynamic of \coc and
\eoc games is equivalent and so will be any quantity computed using
only the state evolution.

The other side of the coin is that, since the state evolution is
independent of the completion-exiting policy, the state itself cannot
entail information about the completion of a strict subset of
collections.

\subsection{Transition probabilities}

Assume to be in a certain state $S(n)\neq\goal$. The $L$ coupons
contained in the lot causing the $(n+1)$-th burst can be partitioned
into $P+1$ subsets. The $l$-th subset has cardinality $R_l$ and
contains the coupons for which exactly $l$ players have a coupon with
the same label. In full analogy with what happens to the state
components, we must have $0\le R_l\le \min\{L,S_l(n)\}$ and
$\sum_{l=0}^P R_l=L$.

There are $\binom{M}{L}$ equally probable lots. The number of lots in
which $R_0$ coupons have a label from $S_0(n)$ possible labels, $R_1$
coupons have a label from $S_1(n)$ possible labels, and so on, is
$\prod_{k=0}^P \binom{S_k(n)}{R_k}$.  Therefore, the probability that
the partition $R_0, R_1,\dots,R_P$ applies to an incoming lot is
\begin{equation}
\label{eq:pR}
\binom{M}{L}^{-1}
\prod_{k=0}^P \binom{S_k(n)}{R_k}
\end{equation}

Given such a partition it is easy to reason as follows. The $R_P$
coupons in the $P$-th subset are discarded since no player needs them.

Each of the $R_{P-k}$ coupons in the $(P-k)$-th subset (for
$k=1,\dots,P-1$), can be assigned to $k$ players. Given the perfect
cooperation between players all those coupons will be assigned.

Hence, since the $R_{P-k}$ coupons are all different, the number of
labels for which $P-k+1$ players hold a coupon ($S_{P-k+1}(n)$)
increases by $R_{P-k}$ while the number of labels for which $P-k$
players hold a coupon ($S_{P-k}(n)$) decreases by the same amount.

Putting all together, the transitions from $S(n)$ caused by a lot of
coupons partitioned into subsets of cardinality $R_0,R_1,\dots,R_P$
leads to an $S(n+1)$ such that
\[
S_k(n+1)=S_k(n)+
\begin{cases}
R_{P-1} & \text{if $k=P$}\\
R_{k-1}-R_k & \text{if $0<k<P$}\\
-R_0 & \text{if $k=0$}
\end{cases}
\]

Conversely, if we know $S(n)$ and $S(n+1)$ we also know that the lot
of coupons causing this transition was partitioned into $P+1$ subsets
with cardinalities
\begin{eqnarray*}
\label{Wdef}
\lefteqn{R_k=}\\
&=&\begin{cases}
\displaystyle \sum_{j=k+1}^P S_j(n+1)-S_j(n) & \text{if $0\le k <P$}\\
\begin{array}{l}
\displaystyle L-\sum_{j=0}^{P-1}R_j=\\
\hspace{1cm}
\displaystyle L-\sum_{j=1}^P j(S_j(n+1)-S_j(n))
\end{array} & \text{if $k=P$}
\end{cases}\\
&=&W(S(n+1)-S(n))
\end{eqnarray*}

\noindent where the affine function $W$ is defined between
$(P+1)$-dimensional vectors.

From the expression of $W$ and from the constraints on the $R_k$ we
have that feasible transitions are those for which
\begin{equation}
\label{cond1}
0\le \sum_{j=k+1}^P S_j(n+1)-S_j(n)\le \min\{L,S_k(n)\}
\end{equation}

\noindent for $0\le k<P$ and

\begin{equation}
\label{cond2}
0\le L-\sum_{j=1}^P j(S_j(n+1)-S_j(n))\le \min\{L,S_P(n)\}
\end{equation}

For this reason, to arrive at a synthetic writing for $T_{S'',S'}$ it
is convenient to define the function
\begin{eqnarray*}
\lefteqn{\phi(S'',S')=}\\
&=&\begin{cases}
1& \text{if
$\begin{array}{l}
\displaystyle 0\le \sum_{j=k+1}^P S''_j-S'_j\le \min\{L,S'_k\}\\
\displaystyle
0\le L-\sum_{j=1}^P j(S''_j-S'_j)\le \min\{L,S'_P\}
\end{array}$}\\
0 & \text{otherwise}
\end{cases}
\end{eqnarray*}

\noindent that evaluates to $1$ if the transition from $S(n)=S'$ to
$S(n+1)=S''$ is feasible and to zero otherwise.

With this, we may recall \eqref{eq:pR} to write
\[
T_{S'',S'}=\phi(S'',S')
\binom{M}{L}^{-1}
\prod_{k=0}^P \binom{S'_k}{W_k(S''-S')}
\]

\subsubsection{Matrix representation}

The probabilities $T_{S'',S'}$ of all possible transitions can be
arranged into a transition matrix (that, with a slight abuse of
notation, we will also indicate with $T$ indexed by a pair of
integers) by means of a $P$-dimensional embedding of integer tuples
into single integers.

This mapping must disregard one of the component of the state tuple
that is redundant. We choose not to consider $S_0$.

This decided, we want to map the set of integer $P$-tuples
$(S_P,\dots,S_1)$ such that $S_l\ge 0$ and $\sum_{l=1}^{P}S_l\le M$
into the set of integers $\{0,1,\dots,\mu_P(M)-1\}$.

We indicate with $Q_P(S_P,\dots,S_1)$ such a mapping in which the
function $Q_P$ is defined by the following recursion
\begin{eqnarray*}
\lefteqn{Q_P(x_1,\dots,x_P)=}\\
&=&\begin{cases}
x_1 & \text{if $P=1$}\\
\displaystyle
\mu_P\left(\sum_{k=1}^{P}x_k-1\right)+Q_{P-1}(x_1,\dots,x_{P-1}) & \text{if $P>1$}
\end{cases}
\end{eqnarray*}

\noindent that can be unrolled to give $Q_P(S_P,\dots,S_1)$ an
explicit expression
\begin{eqnarray*}
\lefteqn{Q_P(S_P,\dots,S_1)=}\\
&=&\mu_P\left(\sum_{k=1}^P S_k-1\right)+
\mu_{P-1}\left(\sum_{k=2}^P S_k-1\right)+\dots+S_P\\
&=& \sum_{j=0}^{P-1}
\mu_{P-j}\left(
\sum_{k=j+1}^P S_k-1
\right)
\end{eqnarray*}

Let us prove that $Q_P$ is a bijection by induction on $P$. For $P=1$
the fact is trivial. Assume then that $P>1$ and that $Q_{P-1}$ is a
bijection.

Given an integer $q\in\{0,\dots,\mu_P(M)-1\}$ we may compute the
$x_1,\dots,x_P$ such that $q=Q_P(x_1,\dots,x_P)$ in few steps.

First, we derive the value of $s=\sum_{k=1}^{P}x_k$ directly from
$q=Q_P(x_1,\dots,x_P)=\mu_P(s-1)+Q_{P-1}(x_1,\dots,x_{P-1})$.  In
fact, if $s$ were given, the minimum for $q$ would be $\mu_P(s-1)$
(obtained for $x_P=s$ and $x_l=0$ for $l<P$) while its maximum would
be $\mu_P(s-1)+\mu_{P-1}(s)-1$ (obtained for $x_1=s$ and $x_l=0$ for
$l>1$). Yet, since by the well-known property of the binomial
coefficients $\mu_P(s-1)+\mu_{P-1}(s)=\mu_P(s)$, we conclude that
$q\in\{\mu_P(s-1),\dots,\mu_P(s)-1\}$. Since for $s=0,1,\dots$ those
ranges are disjoint, $s$ can be directly inferred from $q$.

Once that $s$ is known, we may compute
$Q_{P-1}(x_1,\dots,x_{P-1})=q-\mu_P(s-1)$ and, since $Q_{P-1}$ is a
bijection we also know $x_1,\dots,x_{P-1}$. From these we finally get
$x_P=s-\sum_{k=1}^{P-1}x_k$.

Beyond being invertible, we also have that $Q_P(0,\dots,0)=0$ and
$Q_P(\goal)=\mu_P(M)-1$ so that it is also a bijection between the
possible states $S\neq\goal$ and the integers
$\{0,\dots,\mu_P(M)-2\}$. Hence, we may define the
$(\mu_P(M)-1)\times(\mu_P(M)-1)$ transition matrix
\[
T_{Q_P(S''),Q_P(S')}=\Pr\{S(n+1)=S''|S(n)=S'\}
\]

\noindent for all $S',S''\neq\goal$.

Note that \eqref{cond1} guarantees that, for any feasible transition
we have $\sum_{j=k+1}^P S_j(n+1)\ge \sum_{j=k+1}^P S_j(n)$ for $0\le
k<P$ so that $Q_P(S(n+1))\ge Q_P(S(n))$. Since the entries of
$T_{Q_P(S''),Q_P(S')}$ corresponding to non-feasible transitions are
null, the matrix $T$ is lower triangular.

From now on, the same embedding used to arrange transition
probabilities in the matrix $T$ will be used also to arrange
first-order probabilities $p_{S}(n)$ in a one-dimensional array of
real numbers also indicated by $p$ and defined as
\[
p_{Q_P(S')}(n)=\Pr\{S(n)=S'\}
\]

In an analogous way, we will adopt the general convention of
considering any integer index $q\in\{0,\dots,\mu_P(M)-2\}$ as the
representative of a state $S=Q_P^{-1}(q)\neq\goal$ to allow
calculations to be expressed in terms of matrix operations.

With this, for example, the relationship between the first-order
probabilities after $n$ burst and the first-order probabilities after
$n+1$ bursts can be rewritten exploiting standard matrix product as
$p(n+1)=T p(n)=T^n p(0)$.

To exemplify the construction of the transition matrix $T$ assume
first that $P=1$. In this case the state is an integer scalar simply
accounting for the number of coupons in the collection of the unique
player and the transition matrix can be written straightforwardly.

For $P=1$, $M=3$ and $L=1$ the matrix $T$ is
\[
T=\left(
\begin{array}{lll}
 0 & 0 & 0 \\
 1 & \nicefrac{1}{3} & 0 \\
 0 & \nicefrac{2}{3} & \nicefrac{2}{3}
\end{array}
\right)
\]

\noindent in which the first null row (and thus null eigenvalue)
corresponds to the fact that the $0$-th state cannot be reached from
any other state. Note that, as declared before, we are not interested
into transitions to the final state that do not appear in the matrix.

If we increase $L$ to 2 the transition matrix reflects this by
featuring a further null row corresponding to the new unreachable
state in which only one coupon entered the game. The matrix is then
\[
T=\left(
\begin{array}{lll}
 0 & 0 & 0 \\
 0 & 0 & 0 \\
 1 & \nicefrac{2}{3} & \nicefrac{1}{3}
\end{array}
\right)
\]

When $P=2$ the state is the integer pair $(S_2,S_1)$ whose feasible
values can be mapped to the set of integers
$\{0,1,\dots,(M+1)(M+2)/2-1\}$. Since transitions to the final state
$(M,0)$ are not taken into account, the resulting matrix is an
$[(M+1)(M+2)/2-1] \times [(M+1)(M+2)/2-1]$.

\begin{figure*}[t!]
\begin{center}
\begin{minipage}[c]{0.4\columnwidth}
\includegraphics[width=\columnwidth]{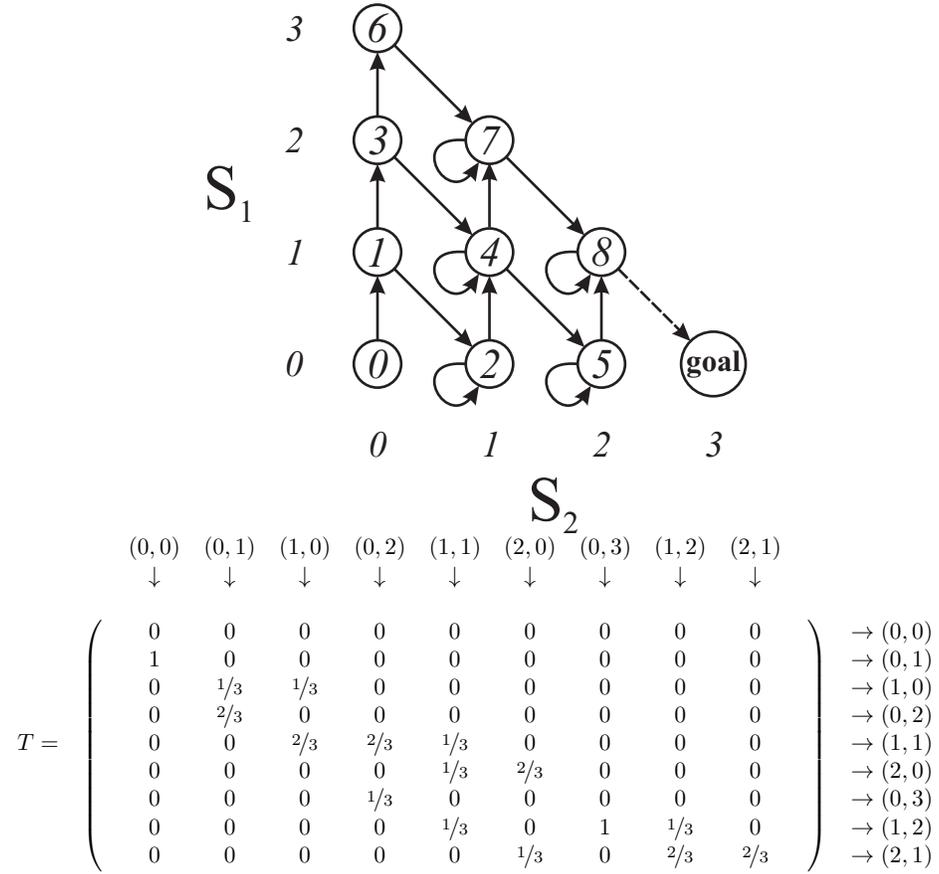}
\end{minipage}
\ \ \ \ \ \ \ \ 
\begin{minipage}[c]{0.7\columnwidth}
\resizebox{\columnwidth}{!}{
$
\begin{array}{llcccccccccrr}
&& (0,0) & (0,1) & (1,0) & (0,2) & (1,1) & (2,0) & (0,3) & (1,2) &
  (2,1) \\
&& \downarrow & \downarrow & \downarrow & \downarrow & \downarrow &
  \downarrow & \downarrow & \downarrow & \downarrow \\
\\
&\multirow{9}{*}{$\left(\rule{0cm}{2cm}\right.$}
& 0 & 0 & 0 & 0 & 0 & 0 & 0 & 0 & 0 & 
\multirow{9}{*}{$\left.\rule{0cm}{2cm}\right)$} & \rightarrow(0,0)\\
&& 1 & 0 & 0 & 0 & 0 & 0 & 0 & 0 & 0 & & \rightarrow(0,1)\\
&& 0 & \nicefrac{1}{3} & \nicefrac{1}{3} & 0 & 0 & 0 & 0 & 0 & 0 & & \rightarrow(1,0)\\
&& 0 & \nicefrac{2}{3} & 0 & 0 & 0 & 0 & 0 & 0 & 0 & & \rightarrow(0,2)\\
T=&& 0 & 0 & \nicefrac{2}{3} & \nicefrac{2}{3} & \nicefrac{1}{3} & 0 & 0 & 0 & 0 & & \rightarrow(1,1)\\
&& 0 & 0 & 0 & 0 & \nicefrac{1}{3} & \nicefrac{2}{3} & 0 & 0 & 0 & & \rightarrow(2,0)\\
&& 0 & 0 & 0 & \nicefrac{1}{3} & 0 & 0 & 0 & 0 & 0 & & \rightarrow(0,3)\\
&& 0 & 0 & 0 & 0 & \nicefrac{1}{3} & 0 & 1 & \nicefrac{1}{3} & 0 & & \rightarrow(1,2)\\
&& 0 & 0 & 0 & 0 & 0 & \nicefrac{1}{3} & 0 & \nicefrac{2}{3} &
\nicefrac{2}{3} & & \rightarrow(2,1)
\end{array}
$
}
\end{minipage}
\end{center}
\caption{\label{cantor231}State embedding and transitions matrix for $P=2$, $M=3$, $L=1$.}
\end{figure*}

For $P=2$, $M=3$ and $L=1$, Figure \ref{cantor231}-left shows how the
two dimensional states different from \goal are mapped into the
integers $\{0,1,2,3,4,5,6,7,8\}$ along with the feasible transitions
between the resulting integer states.

Probabilities for those transitions with the exception of the dashed
one are reported in Figure \ref{cantor231}-right in the form of the
transition matrix $T$.

\begin{figure*}[t!]
\begin{center}
\begin{minipage}[c]{0.4\columnwidth}
\includegraphics[width=\columnwidth]{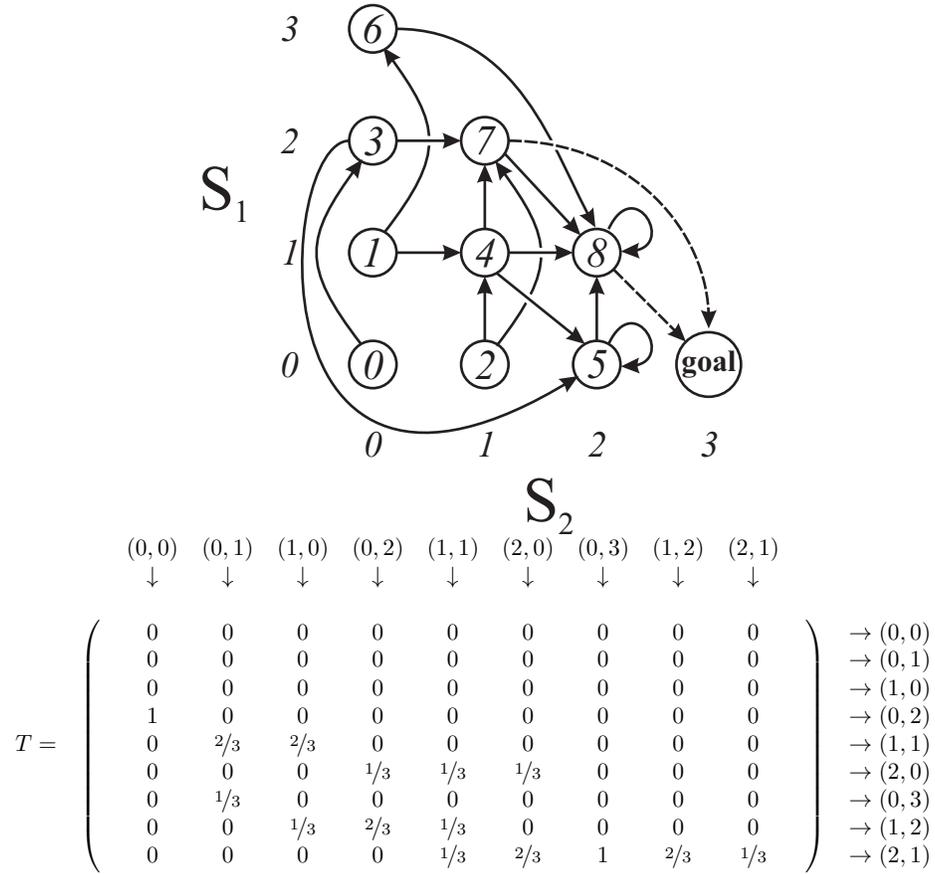}
\end{minipage}
\ \ \ \ \ \ \ \ 
\begin{minipage}[c]{0.7\columnwidth}
\resizebox{\columnwidth}{!}{
$
\begin{array}{llcccccccccrr}
&& (0,0) & (0,1) & (1,0) & (0,2) & (1,1) & (2,0) & (0,3) & (1,2) &
  (2,1) \\
&& \downarrow & \downarrow & \downarrow & \downarrow & \downarrow &
  \downarrow & \downarrow & \downarrow & \downarrow \\
\\
&\multirow{9}{*}{$\left(\rule{0cm}{2cm}\right.$}
& 0 & 0 & 0 & 0 & 0 & 0 & 0 & 0 & 0 & 
\multirow{9}{*}{$\left.\rule{0cm}{2cm}\right)$} & \rightarrow(0,0)\\
&& 0 & 0 & 0 & 0 & 0 & 0 & 0 & 0 & 0 & & \rightarrow(0,1)\\
&& 0 & 0 & 0 & 0 & 0 & 0 & 0 & 0 & 0 & & \rightarrow(1,0)\\
&& 1 & 0 & 0 & 0 & 0 & 0 & 0 & 0 & 0 & & \rightarrow(0,2)\\
T=&& 0 & \nicefrac{2}{3} & \nicefrac{2}{3} & 0 & 0 & 0 & 0 & 0 & 0 & & \rightarrow(1,1) \\
&& 0 & 0 & 0 & \nicefrac{1}{3} & \nicefrac{1}{3} & \nicefrac{1}{3} & 0 & 0 & 0 & & \rightarrow(2,0)\\
&& 0 & \nicefrac{1}{3} & 0 & 0 & 0 & 0 & 0 & 0 & 0 & & \rightarrow(0,3)\\
&& 0 & 0 & \nicefrac{1}{3} & \nicefrac{2}{3} & \nicefrac{1}{3} & 0 & 0 & 0 & 0  & & \rightarrow(1,2)\\
&& 0 & 0 & 0 & 0 & \nicefrac{1}{3} & \nicefrac{2}{3} & 1 & \nicefrac{2}{3} & \nicefrac{1}{3}  & & \rightarrow(2,1)
\end{array}
$
}
\end{minipage}
\end{center}
\caption{\label{cantor232}State embedding and transitions matrix for $P=2$, $M=3$, $L=2$.}
\end{figure*}

For $P=2$, $M=3$ and $L=2$, Figure \ref{cantor232}-left shows how the
two-dimensional states different from \goal are mapped into the
integers $\{0,1,2,3,4,5,6,7,8\}$ along with the feasible transitions
between the resulting integer states.

Probabilities for those transitions with the exception of the dashed
ones are reported in Figure \ref{cantor232}-right in the form of the
transition matrix $T$.

Note that all the Markov chains we are dealing with are finite and
feature only ``forward'' transition up to a unique absorbing state
that is the final goal. This class of Markov chains is well-understood
and the following results are obtained by specializing established
methods (see, e.g., \cite{Kemeny_1960}).

\subsection{Computation of $\EBgoal$}

\begin{theorem}
\label{theEBgoal}

If the $(\mu_P(M)-1)\times(\mu_P(M)-1)$ matrix $T$ is such that
\[
T_{Q_P(S''),Q_P(S')}=\phi(S'',S')
\binom{M}{L}^{-1}
\prod_{k=0}^P \binom{S'_k}{W_k(S''-S')}
\]

\noindent and the numbers $\tau_j$ are defined as
\begin{equation}
\label{taudef}
\tau_j=\begin{cases}
\displaystyle
1 & \text{if $j=0$}\\
\displaystyle
\frac{1}{1-T_{j,j}}
\sum_{k=0}^{j-1}T_{j,k}\tau_k
 &
\text{if $j=1,\dots,\mu_P(M)-2$}
\end{cases}
\end{equation}

\noindent then
\[
\EBgoal=\sum_{j=0}^{\mu_P(M)-2}\tau_j
\]

\end{theorem}

\begin{IEEEproof}
If $\bar{n}$ is such that $S(\bar{n})=\goal$ then
\[
\EBgoal=\Exp[\bar{n}]=
\sum_{n=0}^\infty
\Pr\{\bar{n}>n\}
\]

The path is straightforward. At the beginning, no player has a coupon,
i.e., the system is in the state $S(0)=(0,0,\dots,0,M)$ with
probability 1. Since $Q_P(0,\dots,0)=0$ we have $p(0)=\pstart^\top=v$,
where the column vector $v$ remains implicitly defined.

Starting from this initial condition, the system evolves at each burst
and, after $n$ bursts, is characterized by the first-order
probabilities $p(n)=T^n v$.

The probability that $\bar{n}>n$ is the probability that, after $n$
bursts, the system is in any state but \goal, i.e.,
$\Pr\{\bar{n}>n\}=\sum_{j=0}^{\mu_P(M)-2} p_j(n)= \pcollecttop p(n)=u
p(n)$, where the row vector $u$ remains implicitly defined.
\begin{equation}
\label{eq:EBgoal}
\EBgoal=\sum_{n=0}^\infty u T^n v=u \left(I-T\right)^{-1}v
\end{equation}

\noindent where $I$ is the identity matrix. Note that the inversion of
$(I-T)$ is possible since $T$ is lower triangular and thus exhibits
its eigenvalues on the diagonal, and since such eigenvalues (that are
the probabilities that none of the $L$ coupons of the incoming lot can
be assigned to any player) are less than one if the common goals has
not been achieved yet.

Note finally that $I-T$ is also lower triangular. Hence, the
above expression can be easily expanded component-wise to realize that,
by defining the sequence of $\tau_j$ as in \eqref{taudef}, we get the
thesis.
\end{IEEEproof}

\begin{figure*}[t!]
\begin{center}
\psfragscanon
\psfrag{EBgoal}{$\frac{\EBgoal L}{P}$}
\psfrag{M}[][][0.8]{$M$}
\includegraphics[width=0.7\textwidth]{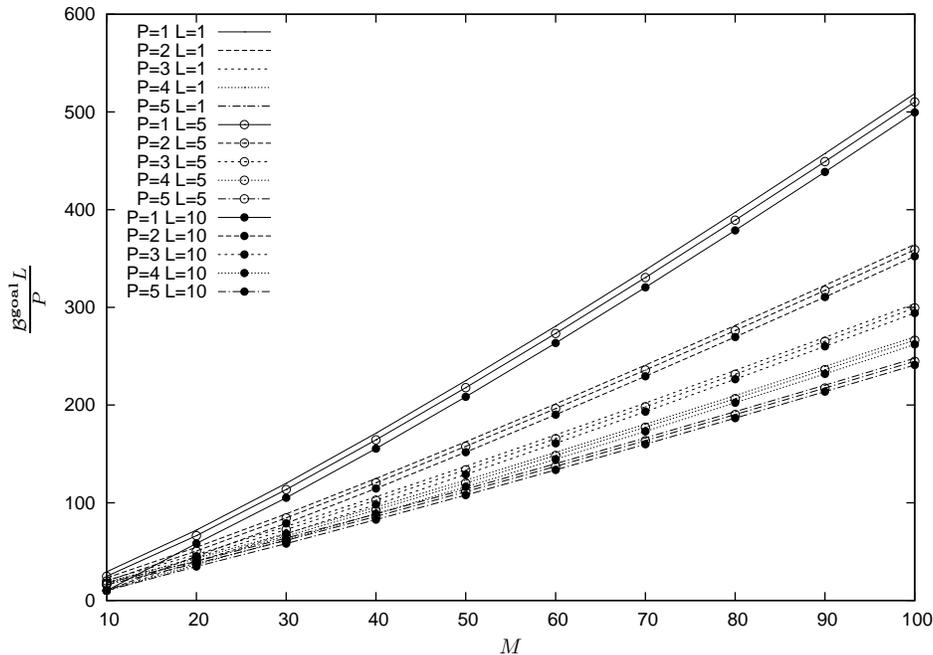}
\psfragscanoff
\end{center}
\caption{\label{EBgoalx}Normalized performance by varying $M$ for different $P$ and $L$.}
\end{figure*}

In Figure \ref{EBgoalx} we report $\frac{\EBgoal L}{P}$ against $M$
for different values of $P$ and $L$. The adoption of a normalized
quantity that measures performance in terms of the average number of
coupons received directly by each player allows to compare different
configurations.

Besides the general and intuitive trend increasing with $M$, the plots
reveal two causes of improvement (i.e., decrease of $\EBgoal$).

The first is the availability of lots with $L>1$. This introduces a
correlation between coupon appearance that benefits the collection
since it reduces the number of duplicates.

The second is the cooperation between players (i.e., the fact that
$P>1$) that introduces additional sources of coupons (the other $P-1$
players) that may contribute to the completion of the collection.

Note how this second effect is, in this plot, much more evident than
the first, to the extent that the case $L=1$ is always a very good
upper bound on $\EBgoal$ computed for other values of $L$.

Actually, this is due to the fact that, for most of the plotted
configurations, $\frac{L}{M}\ll 1$ thus reducing the effect of coupon
correlation in lots.

As a final remark, from \eqref{taudef} we get that the computation
of the $\tau_j$ may greatly benefit from the fact that $T$ is, in
general, quite sparse.

In fact, the number of feasible transitions from the generic
intermediate state $(S_P,S_{P-1},\dots,S_0)$ is equal to the number of
ways in which the integer $L$ can be decomposed as the sum of $P$
integers $0\le R_l\le \min\{L,S_l\}$. Regardless of the state, this
number can be easily bounded from above disregarding the constraint
$R_l\le S_l$.

Following the same path that led us to compute the total number of
feasible state vectors, we obtain that not more than $\binom{L+P}{P}$
transitions are feasible from each intermediate state.

Since there are $\binom{M+P}{P}-1$ states different from the final
goal, the number of non-vanishing terms considered in the
computation of the $\tau_j$ in \eqref{taudef} is not larger than
$\binom{L+P}{P}\left(\binom{M+P}{P}-1\right)$.

\subsection{Computation of $\EBgoal$ for $P=1$}

As noted before, in this case $\mu_1(M)=M+1$ since the state of the
system is the degree of filling of the collection of the unique
player: from 0 coupons to one coupon for each of $M-1$ labels before
achieving \goal.

In this case
\[
T_{j,k}=
\begin{cases}
\binom{M}{L}^{-1}\binom{M-k}{j-k}\binom{k}{L-j+k} &
\text{if $j-k\le L$}\\
0 & \text{otherwise}
\end{cases}
\]

The matrix $T$ can be written as $T=E \Lambda E$ where $\Lambda$ is a
diagonal matrix with eigenvalues
\[
\Lambda_{j,j}=T_{j,j}=\binom{M}{L}^{-1}\displaystyle\binom{j}{L}
\]
\noindent for $j=0,\dots,M-1$ and where
\[
E_{j,k}=\binom{M-k}{M-j}(-1)^{M-j}
\]

\noindent for $j,k=0,\dots,M-1$.

The matrix $E$ is independent of $L$ and has the noteworthy property
of being lower triangular and such that $E^{-1}=E$ or
$EE=I$.

Based on this, we may compute the terms $\tau_j$ in \eqref{taudef}
that are the entries of the first column of
$\left(I-T\right)^{-1}=E(I-\Lambda)^{-1}E$, i.e.
\begin{eqnarray*}
\tau_j&=&\sum_{k=0}^j E_{j,k}\frac{1}{1-T_{k,k}} E_{k,0}\\
&=&\sum_{k=0}^j \binom{M-k}{M-j}(-1)^{M-j}\frac{\binom{M}{L}}{\binom{M}{L}-\binom{k}{L}} \binom{M}{M-k}(-1)^{M-k}\\
&=&\binom{M}{L}\binom{M}{j}\sum_{k=0}^j \binom{j}{k}\frac{(-1)^{j+k}}{\binom{M}{L}-\binom{k}{L}}
\end{eqnarray*}

Plugging this into \eqref{eq:EBgoal} we get
\begin{eqnarray*}
\lefteqn{\EBgoal=}\\
&=&\binom{M}{L}\sum_{j=0}^{M-1}
\binom{M}{j}\sum_{k=0}^j \binom{j}{k}\frac{(-1)^{j+k}}{\binom{M}{L}-\binom{k}{L}}\\
&=&
\binom{M}{L}
\sum_{k=0}^{M-1}
\frac{1}{\binom{M}{L}-\binom{k}{L}}
\sum_{j=k}^{M-1}
\binom{M}{j}\binom{j}{k}(-1)^{j+k}\\
&=&
\binom{M}{L}
\sum_{k=0}^{M-1}
\frac{(-1)^{k+M-1}}{\binom{M}{L}-\binom{k}{L}}
\binom{M}{k}
\end{eqnarray*}

This expression is a special case of equation (5) in \cite{Adler_2001}
for equally probable subsets of $L$ out of $M$ labels.

Moreover, in the special case $L=1$, $\tau_j$ turns out to be
$\tau_j=\frac{M}{M-j}$ thus reproducing the classical
$\EBgoal=M\sum_{j=1}^{M}j^{-1}$ whose $M\log M$ asymptotic trend is
well-known.

\subsection{Computation of $\EBfirst$ for $L=1$}

The computation of this quantity hinges on counting how many coupon
assignments are subsumed by each state $S$ and distinguishing in how
many of them at least one player has finished her collection. This can
be done straightforwardly in the $L=1$ case that is the one addressed
here.

Assume now that $\bar{n}$ is the number of bursts needed by the first
player to complete her collection. As before, the average of this
quantity can be written starting from its complementary cumulative
distribution function as
\[
\EBfirst=\Exp[\bar{n}]=
\sum_{n=0}^\infty
\Pr\{\bar{n}>n\}
\]

In this case $\Pr\{\bar{n}>n\}$ is the probability that after $n$
burst no player has completed her collection and we may expand
\[
\Pr\{\bar{n}>n\}=\sum_{S'\neq\goal} \Pr\{\bar{n}>n|S(n)=S'\}\Pr\{S(n)=S'\}
\]

Note that $S'$ spans all the states with the exception of $\goal$
since the common goal implies that all players have completed their
collections.

To compute $\Pr\{\bar{n}>n|S(n)=S'\}$ assume that
$S'=(S_P,\dots,S_0)$. Corresponding to that state, the $\sum_{j=1}^P j
S_j$ coupons that are in the game, may be assigned to the $P$ players
in
\[
U_P(S_P,\dots,S_0)=\binom{M}{S_P,\dots,S_0}\prod_{j=0}^P \binom{P}{j}^{S_j}
\]

\noindent ways, that are all equally probable.

Among those assignments, there are $V$ in which no player has completed
her collection. This number $V$ is the difference between $U$ and
number of assignments in which exactly $j$ players have completed
their collection, for $j=1,...,P$.

Note now that, if $S_0>0$ then no player may have completed her
collection. In general, if $S_j>0$ then not more than $j$ players may
have completed their collection.

Assuming that it is possible, the number of assignments in which
exactly $j$ players have completed their collections is the number of
choices of $j$ players out of $P$ (i.e., $\binom{P}{j}$) times the
number of assignments of the remaining $\sum_{k=1}^P k S_k-Mj$ coupons
to the residual $P-j$ players such that none of them have completed their
collections. To count these assignments note that since we drop $j$
players that have a coupon for each possible label, if $S_k$ was the
number of labels for which $k$-players had a coupon before, now only
$k-j$ out of the remaining $P-j$ have a coupon for those labels.

Hence we may recursively write
\begin{eqnarray*}
\lefteqn{V_P(S_P,\dots,S_0)=U_P(S_P,\dots,S_0)+}\\
&&
\displaystyle
\hspace{1cm}-\sum_{j=1}^{\max\{k|S_0=\dots=S_{k-1}=0\}}
\binom{P}{j}
V_{P-j}(S_P,\dots,S_j)
\end{eqnarray*}

\noindent to yield
\[
\Pr\{\bar{n}>n|S(n)=S'\}=
\frac{V_P(S')}{U_P(S')}
\]

Assuming to align all the above conditioned probabilities in the
$(\mu_P(M)-1)$-dimensional row vector $\nu$ such that
$\nu_j=\Pr\{\bar{n}>n|S(n)=Q_P^{-1}(j)\}$, a path identical to what was
followed for $\EBgoal$ leads to
\[
\EBfirst=\nu \left(I-T\right)^{-1} v
\]

Relying on the previous \eqref{taudef} we may finally write
\[
\EBfirst=\sum_{j=0}^{\mu_P(M)-2}\nu_j\tau_j
\]

\begin{figure*}[t!]
\begin{center}
\psfragscanon
\psfrag{EBs}{$\frac{\EBgoal}{P}, \frac{\EBthis}{P}, \frac{\EBfirst}{P}$}
\psfrag{M}[][][0.8]{$M$}
\psfrag{P=4 Bthis}[][][0.7]{$P=4\,\EBthis$}
\psfrag{P=4 Bgoal}[][][0.7]{$P=4\,\EBgoal$}
\psfrag{P=4 Bfirst}[][][0.7]{$P=4\,\EBfirst$}
\psfrag{P=2 Bthis}[][][0.7]{$P=2\,\EBthis$}
\psfrag{P=2 Bgoal}[][][0.7]{$P=2\,\EBgoal$}
\psfrag{P=2 Bfirst}[][][0.7]{$P=2\,\EBfirst$}
\psfrag{P=1 Bgoal, Bfirst, Bthis}[][][0.7]{$P=1\,\EBgoal,\EBthis,\EBfirst$}
\includegraphics[width=0.7\textwidth]{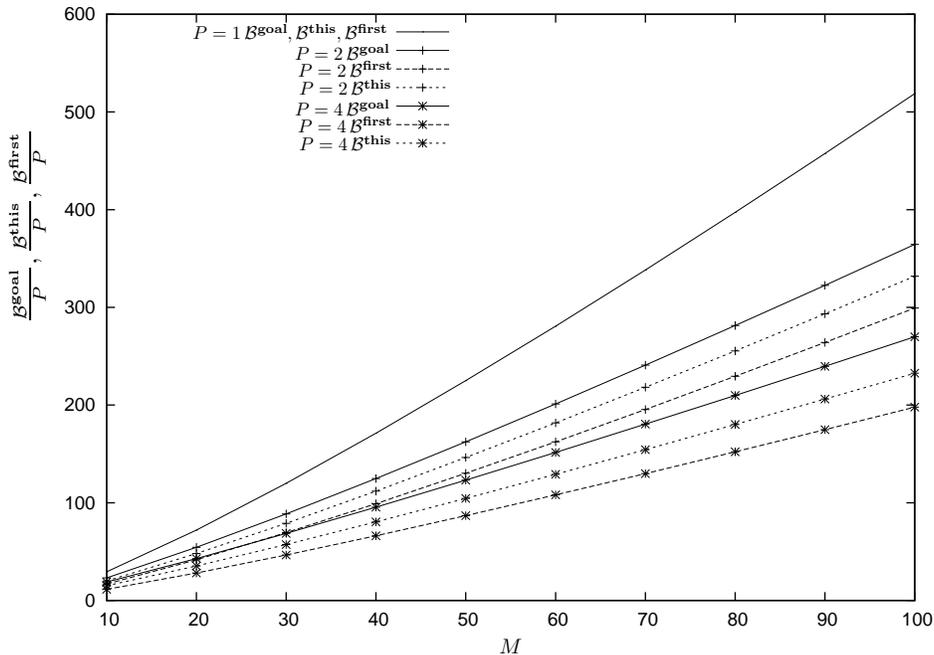}
\psfragscanoff
\end{center}
\caption{\label{EBsx}Global performance compared with performance relevant to \eoc games by varying $M$ for different $P$.}
\end{figure*}

\subsection{Computation of $\EBthis$ for $L=1$}

As before, the derivation hinges on counting how many coupon
assignments are subsumed by each state $S$ and distinguishing in how
many of them the chosen player has finished her collection. This can
be done straightforwardly in the $L=1$ case that is the one addressed
here.

Assume now that $\bar{n}$ is the number of bursts needed by a
specified player to complete her collection. As before, the average of
this new quantity can be written starting from its complementary
cumulative distribution function as
\[
\EBthis=\Exp[\bar{n}]=
\sum_{n=0}^\infty
\Pr\{\bar{n}>n\}
\]

In this case $\Pr\{\bar{n}>n\}$ is the probability that after $n$
burst the chosen player has not completed her collection and we may
expand
\[
\Pr\{\bar{n}>n\}=\sum_{S'\neq\goal} \Pr\{\bar{n}>n|S(n)=S'\}\Pr\{S(n)=S'\}
\]

Note that $S'$ spans all the states with the exception of $\goal$
since the common goal implies that all players have completed their
collections and that
\[
\Pr\{\bar{n}>n|S(n)=S'\}=1-\prod_{k=0}^P \left(\frac{k}{P}\right)^{S'_k}
\]

Assuming to align all the above conditioned probabilities in the
$(\mu_P(M)-1)$-dimensional row vector $\xi$ such that
$\xi_j=\Pr\{\bar{n}>n|S(n)=Q_P^{-1}(j)\}$, a path identical to what was
followed for $\EBgoal$ leads to
\[
\EBthis=\xi \left(I-T\right)^{-1} v
\]

Relying on the previous \eqref{taudef} we may finally write
\[
\EBthis=\sum_{j=0}^{\mu_P(M)-2}\xi_j\tau_j
\]

\subsection{Details of performance for $L=1$}

In Figure \ref{EBsx} we report $\frac{\EBgoal}{P}$,
$\frac{\EBthis}{P}$, and $\frac{\EBfirst}{P}$ against $M$ for
different values of $P$ and for $L=1$. As before we measure
performance in terms of the average number of coupons received
directly by each player.

As expected, we always have $\EBfirst\le \EBthis \le \EBgoal$.
For $P=1$, $\EBthis=\EBfirst=\EBgoal$, while simple symmetry implies that
$\EBthis=(\EBfirst+\EBgoal)/2$ for $P=2$.
For $P=4$ we still have $\EBthis\approx(\EBfirst+\EBgoal)/2$.

Finally, besides the general and intuitive trend increasing with $M$,
the plots reveal that improvement due to cooperation applies to all
performance figures.

\section{Conclusions}

This work deals with the statistical characterization of multiplayer
coupon collector's games that are a generalization of classical coupon
collector's games with perspective applications in several Information
Technology fields.

What is addressed is the combination of benefits and costs due to the
possibility of cooperation between $P$ players by means exchanging of
coupons that enter the game in lots each of $L$ different units. The
local goal of each player is to complete her collection of $M$
distinct coupons. The global goal is the completion of all
collections.

Such an exchange process is regulated by a protocol entailing offer,
request and transfer phases. Two playing mechanisms are analyzed: one
in which players who complete their collection exit the game (\eoc
games), and one in which they remain active and contribute to the
overall exchanging (\coc games).

The average cost of offer ($\EOC$), request ($\ERC$) and transfer
($\ETC$) phases is computed in analytical terms all yielding very
simple closed form expressions. The quantities $\ERC$ and $\ETC$ turn
out to be independent of $L$ and of the statistics of lot drawing.

Costs are larger for \coc games than for \eoc ones.

As far as performance is concerned, an analytical form is given for
the average number of activity bursts needed by the first player to
complete her collection ($\EBfirst$), a chosen player to complete the
collection ($\EBthis$) and by all players to complete their
collections ($\EBgoal$).

In this case, the equivalence between \coc and \eoc games is proved as
far as $\EBfirst$ and $\EBgoal$ are concerned, and when $L=1$ for
$\EBthis$. Computations of this merit figures quantifies the
effectiveness of cooperation under different points of view.

\end{document}